\documentclass[conference]{IEEEtran}
\IEEEoverridecommandlockouts
\usepackage{makecell}
\usepackage{indentfirst}
\usepackage{multirow}
\usepackage{cite}
\usepackage{amsmath,amssymb,amsfonts}
\usepackage{algorithmic}
\usepackage{graphicx}
\usepackage{textcomp}
\usepackage{xcolor}
\usepackage{tabularx}
\usepackage{makecell,multirow,diagbox}
\usepackage{amssymb}
\usepackage{listings}
\usepackage{subfigure} 
\usepackage{graphicx}
\usepackage{hyperref}
\usepackage{booktabs}
\usepackage{color}
\definecolor{dkgreen}{rgb}{0,0.6,0}
\definecolor{gray}{rgb}{0.5,0.5,0.5}
\definecolor{mauve}{rgb}{0.58,0,0.82}
\def\BibTeX{{\rm B\kern-.05em{\sc i\kern-.025em b}\kern-.08em
    T\kern-.1667em\lower.7ex\hbox{E}\kern-.125emX}}

\begin{document}

\title{BLESER: Bug Localization Based on Enhanced Semantic Retrieval}

\author{\IEEEauthorblockN{Weiqin Zou $^{1,2}$, Enming Li$^{3}$, Chunrong Fang$^{3}$\thanks{Chunrong Fang is the corresponding author.}}
\IEEEauthorblockA{\textit{$^{1}$College of Computer Science and Technology, Nanjing University of Aeronautics and Astronautics, China } \\ \textit{$^{2}$ Key Laboratory of Safety-Critical Software (NUAA), Ministry of Industry and Information Technology, China } \\ \textit{$^{3}$ State Key Laboratory for Novel Software Technology, Nanjing University, China } \\
Email:~weiqin@nuaa.edu.cn~mf1932088@smail.nju.edu.cn~fangchunrong@nju.edu.cn}}

\maketitle

\begin{abstract}
Bug localization techniques play an important role in software quality assurance, among which, static bug localization techniques that locate bugs at method granularity have gained much attention from both researchers and practitioners.
For a static method-level bug localization technique, a key but challenging step is to fully retrieve the semantics of methods and bug reports.
Currently, existing studies mainly use the same bag-of-word space to represent the semantics of methods and bug reports without considering structure information of methods and textual contexts of bug reports, which largely and negatively affects bug localization performance.

To address this problem, we develop BLESER, a new bug localization technique based on enhanced semantic retrieval.
Specifically, we use an AST-based code embedding model (capturing code structure better) to retrieve the semantics of methods, and word embedding models (capturing textual contexts better) to represent the semantics of bug reports.
Then, a deep learning model is built on the enhanced semantic representations.
During model building, we compare five typical word embedding models in representing bug reports and try to explore the usefulness of re-sampling strategies and cost-sensitive strategies in handling class imbalance problem.
We evaluate our BLESER on five Java projects from the Defects4J dataset.
We find that:
(1) On the whole, the word embedding model ELMo outperformed the other four models (including word2vec, BERT, etc.) in facilitating bug localization techniques.
(2) Among four strategies aiming at solving class imbalance problems, the strategy ROS (random over-sampling) performed much better than the other three strategies (including random under-sampling, Focal Loss, etc.).
(3) By integrating ELMo and ROS into BLESER, at method-level bug localization, we could achieve MAP of 0.108-0.504, MRR of 0.134-0.510, and Accuracy@1 of 0.125-0.5 on five Defects4J projects.
\end{abstract}


\section{Introduction}
Bug localization techniques play an important role in ensuring the quality of software products and have always been attracting much attention from both the industry and research community \cite{kochhar2016practitioners, zou2018practitioners}.
Existing studies could be broadly divided into two categories, namely \textit{dynamic} and \textit{static} bug localization techniques \cite{zhang2019finelocator}.
Dynamic techniques mainly aim to collect execution profiles of programs under test to locate buggy code, while static techniques mainly aim to retrieve some static features from bug reports, source code, and other artifacts (e.g., commit logs) generated within the process of software development, to locate buggy code.
Different bug localization techniques locate bugs at different source code granularity levels (including file, method, statement, etc.).
In this paper, we mainly focus on static method-level bug localization techniques.
More specifically, we aim to develop a technique that recommends buggy \textit{methods} for a given bug report,
as the method granularity is found to be mostly preferred by software practitioners \cite{kochhar2016practitioners}.

For static method-level bug localization techniques, a key challenge is to fully retrieve the functional semantics of a method (i.e., the functionality of a method) and problem semantics of a bug report (i.e., the exact software problem a bug report describes).
Currently, existing studies mainly used traditional information retrieval techniques to retrieve the semantics of method code and bug reports in the same bag-of-words feature space, i.e., both method code and bug reports are taken as textual content in natural language\cite{lukins2010bug, zhang2016inferring, zhang2019finelocator}.
This is problematic in two aspects.
On the one hand, taking method code as pure text would miss the structure information of a method beyond its lexical terms.
Since statements organized in different ways (e.g., from \texttt{A->B->C to B->A->C}) are very likely to reveal different functional semantics, ignoring the code structure would lead to insufficient semantic retrieval of a method.
On the other hand, the information retrieval techniques (such as LDA (Latent Dirichlet Allocation), VSM (Vector Space Model)) used by previous studies usually fail to deal with the context of lexical terms, which also leads to insufficient semantic representations of bug reports and methods in terms of their textual content.
The above-mentioned problems in semantic retrieval of methods and bug reports largely and negatively affect the performance of bug localization techniques.

In this paper, we proposed a technique called BLESER (Bug Localization based on Enhanced SEmantic Retrieval), to recommend buggy methods for a given bug report.
Unlike previous research, BLESER took full consideration of code structure and textual context while retrieving the semantics of methods and bug reports.
Specifically, we used an abstract syntax tree (AST) based model to retrieve the functional semantics of methods.
The semantics of bug reports were retrieved by state-of-the-art word embedding models which have been proved to be effective in representing textual semantics in various natural language tasks \cite{li2018bridging, ye2016word, zhang2019finelocator}.
Then, we used a deep learning model that leverages two kinds of semantic features of methods and bug reports, to learn unified features from methods and bug reports to automatically locate buggy methods for a given bug report.

Furthermore, although word embedding models are widely used in natural language tasks, they are rarely well studied under the context of bug localization. Hence, we do not have a good reference in deciding which word embedding model to use in BLESER.
To help understand how effective typical word embedding models are in representing textual semantics of bug reports, we did a comprehensive comparison of five classical and state-of-the-art word embedding models in this study, namely word2vec, GloVe, fastText, ELMo, and Bert\footnote{https://github.com/flairNLP/flair}.
What is more, during model building, we found that the training dataset is extremely imbalanced; that is, the number of buggy instances (methods) are much smaller than the number of non-buggy instances for a given bug report.
To avoid the potential threats of imbalanced datasets upon model building, we checked the effectiveness of four imbalanced-class-handling strategies in our experiments, and integrated the most effective strategy into BLESER to fix the imbalanced-class problem.

Our experiments are conducted on the benchmark dataset called Defects4J\footnote{https://github.com/rjust/Defects4J}, which consists of five Java projects.
Defects4J is often used as the dataset to evaluate various bug localization and repair tasks \cite{b2016learning, li2019deepfl, martinez2016astor}.
Through experiments, we find that 
ELMo performed best among five word embedding models in facilitating the performance of bug localization techniques, which means ELMo is a more suitable choice in representing the semantics of bug reports than other models.
Meanwhile, we find that the resampling strategy ROS (random over-sampling) could always help achieve the best localization performance, which means ROS is mostly suitable in handling the class imbalance problem of bug localization.
Our BLESER which integrates ELMo and ROS could achieve MAP of 0.108-0.504, MRR of 0.134-0.510, and Accuracy@1 of 0.125-0.5 on five Defects4J projects.
Our major contributions are as follows.
\begin{itemize}

    \item We proposed a technique called BLESER which takes code structure of methods and textual context of bug reports into account in representing the semantics of methods and bug reports, to locate buggy methods for given bug reports. Experimental results demonstrate the effectiveness of our approach.

    \item We studied the potential of five typical word embedding models in retrieving the semantics of bug reports. We found that ELMo could help obtain the best performance in bug localization.

    \item We explored the effectiveness of four strategies in handling imbalanced-class problems within bug localization prediction models. We found that the strategy random over-sampling was found to be most helpful in locating bugs.
\end{itemize}
The remaining parts are organized as follows.
Section 2 introduces the background. Section 3 describes the framework of our technique.
Section 4 and 5 present the experiment setup and results. Section 6 shows threats to the validity of our study. Section 7 introduces related work. Finally, Section 8 summarizes our findings and presents future work.

\section{Background}
In this section, we mainly introduce two kinds of techniques, namely word embedding and AST-based code embedding, that we used to enhance the semantic retrieval of bug reports and method code, respectively.

\subsection{Word Embedding}
In 1954, Harris proposed a distributed hypothesis that words appearing in the same or similar contexts usually have similar meanings \cite{harris1954distributional}.
Since then, various distributed semantic models (DSMs) are proposed.
In a DSM, each word is represented by an N-dimensional numeric vector.
Words that appear in similar or the same context have similar vector representations.
Most traditional DSMs are mainly based on counting statistics.
In recent years, a series of DSMs based on neural networks are proposed \cite{collobert2008unified, mikolov2013distributed}.
A neural-network-based DSM mainly applies a deep-learning neural network to learn the contexts of textual words and convert each word into a low-dimensional numeric vector.
Such a DSM is called a word embedding model.
A word embedding model could better capture the semantics of word context than those DSMs based on counting statistics; and therefore is more effective in various natural language tasks \cite{chen2016mining, ye2016word, li2018bridging}.

In the area of natural language processing, researchers and practitioners have proposed a series of word embedding models. Some representative models are word2vec \cite{mikolov2013distributed}, ELMo, GloVe\footnote{https://nlp.stanford.edu/projects/glove/}, BERT and fastText\footnote{https://fasttext.cc/}. 
Google developed word2vec in 2013. This tool implements two different word embedding algorithms, including CBOW and Skip-gram.
The CBOW model aims to predict the center word by proving its contextual words; while the Skip-gram model aims to predict the contextual words of a given center word.
Both models use gradient descent to adjust the weights of individual network nodes during model training.
Then for each word, we could obtain its word embedding numeric vector from the built models.
As a traditional word embedding technique, word2vec has been used to help solve some software engineering tasks \cite{chen2016mining, ye2016word}.
One major disadvantage of word2vec is that it cannot distinguish words that may show different meanings under different contexts, as each word only has one corresponding numeric vector.

GloVe makes use of some global statistical information and solves some problems of word co-occurrence models. 
ELMo uses bidirectional LSTM to extract semantic features. It first adopts a language model to obtain the numeric vector of each word, and then dynamically adjusts the word vector by using the word contexts of downstream tasks.
The adjusted word vector can better represent the concrete meaning of a word under a certain context, which further solves the polysemous problem of word2vec. 
One major difference between BERT and ELMo is that BERT adopts Transformer rather than LSTM to extract semantic features of the text.

Unlike above-mentioned word embedding techniques, fastText (proposed by Facebook) is trained on N-gram word-bag-level data rather than on word-level data.
Specifically, fastText represents words as N-gram word bags, and then uses Skip-gram to train a model on these word bags.
Each N-gram subword corresponds to a numeric vector, and a word could be represented as the aggregation of the N-gram vectors of its subwords.
By training on N-gram-based subwords, fastText could capture the semantics of shorter words, rare words and even prefixes and suffixes.

Currently, embedding techniques such as ELMo have not been adopted in the task of method-level bug localization. To understand how useful these embedding models are in helping bug localization, we conduct a comparative analysis of the above-mentioned five typical/state-of-the-art word embedding methods and try to find the most suitable word embedding model for representing the semantics of bug reports.

\subsection{AST-based Code Embedding}
Similar to word embedding models, code embedding models aim to learn the semantics of code snippets and represent them as continuously distributed numeric vectors \cite{alon2019code2vec, zhang2019novel}.
These code vectors could be used to facilitate software tasks such as code clone detection or code naming \cite{white2016deep, alon2019code2vec}.
In recent years, with the rapid development of deep learning algorithms, researchers proposed a diverse code embedding models based on deep learning methods;
and a typical category of such models is AST-based code embedding models, i.e., applying deep learning methods to abstract syntax trees of source code.
An AST is a tree designed to characterize the abstract syntax structure of source code.
Each node of an AST corresponds to a construct in the source code.
Since an AST could fully represent the lexical and grammatical structure of the source code without recording all details of the code (such as punctuation marks), it is widely used in many software engineering techniques/tools.
Some AST-based deep learning code embedding techniques include RvNN, TBCNN, Tree-LSTM, ASTNN and code2vec\cite{alon2019code2vec, mou2016convolutional, socher2011parsing, tai2015improved, zhang2019novel}.
In this paper, we mainly use code2vec to represent the semantics of method code.

code2vec was proposed by Alon et al. in their POPL 2019 paper.
code2vec is a code embedding model that adopts an attention neural network to learn a matching relationship between a bag of path-contexts and a label (i.e., method name).
Within code2vec, a code snippet is first converted into an AST.
Then from the AST, all paths from a leaf node to another leaf node are extracted.
After that, an attention neural network is applied to these paths to obtain their weights and embedding vectors.
Finally, those embedding vectors of all paths are aggregated as a final embedding vector to represent the semantics of a code snippet.
To demonstrate the correctness of code embedding vectors, the authors used the generated code vectors by code2vec to predict method names for given code snippets and achieved impressive results.
This means the code vectors generated by code2vec could well represent the semantics of method code.
Hence, in this paper, we decided to use code2vec to obtain code embedding vectors to represent the semantics of methods while building our method-level bug localization technique.

\section{Methodology}
\begin{figure*}[!htb]
\center
\vspace{-4.4cm}
\includegraphics[scale=0.75]{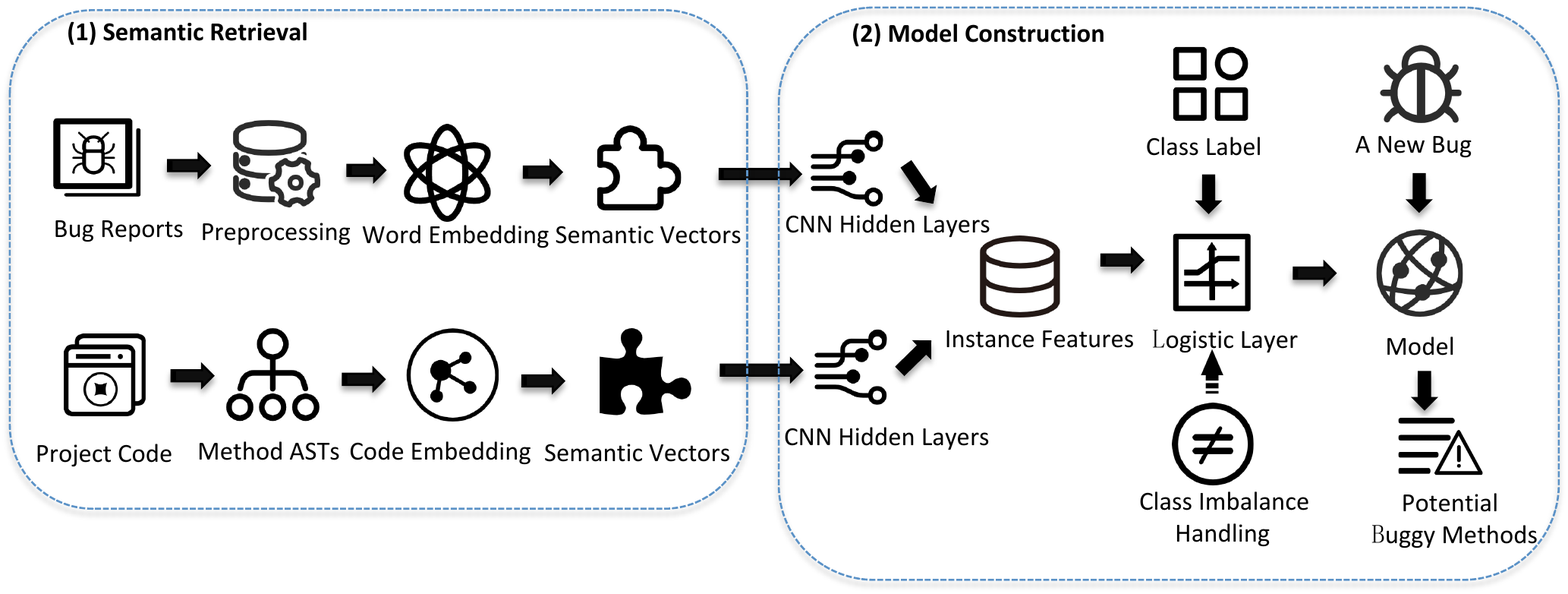}\vspace{-5.8cm}
\caption{The Framework of BLESER.}
\label{fig_framework}
\end{figure*}

Fig. \ref{fig_framework} shows the overall framework of BLESER.
As a method-level bug localization technique, BLESER mainly includes two parts, namely semantic retrieval and model building.
In the semantic retrieval part, we mainly use word embedding models and code embedding models to retrieve the semantic features (i.e., numeric embedding vectors) of bug reports and method code, respectively.
Particularly, we use pre-trained embedding models to perform semantic retrieval.
After we obtain the semantic features of bug reports and methods, we use a neural network to unify the two kinds of semantic feature vectors.
Last, we use logistic regression as the output layer of the neural network, to train a model based on training instances (with unified features as instance features).
The model then could be used to predict potential buggy methods for a new coming bug report.
During model training, we found that the training dataset is extremely imbalanced (i.e., the number of instances from one class is much larger (smaller) than that from another class).
To alleviate the potential problems brought by imbalanced classes, we incorporated relevant strategies handling imbalanced classes into BLESER.   
The details of our method are as follows.

\subsection{Semantic Retrieval}
\textbf{Pre-trained Embedding Models.}
Inspired by the effectiveness of word embedding and code embedding techniques in natural language tasks and other software engineering tasks, we proposed to use these embedding models to improve the semantic retrieval of bug reports and method code.
Generally speaking, a large-scale training dataset is always a prerequisite for building a deep-learning-based word embedding model or code embedding model of good-performance.
Whereas, many software projects (so do our experimental projects) do not have massive bug reports or method code for us to build such an embedding model.
To solve this problem, we decided to use pre-trained models to retrieve the semantics of bug reports and methods in this study.

A pre-trained model is one solution proposed by researchers to the problem that a certain task does not have sufficient data to build its own effective model from the scratch \cite{peters2018deep}.
In other words, rather than build a new model, we could use an available model built upon publicly massive benchmark datasets from the same or relevant domains to resolve our problems.
Currently, pre-trained models have been widely used in image processing and natural language processing tasks \cite{wang2016cnn, mikolov2017advances}.
For example, some image segmentation tasks used pre-trained models built on the public ImageNet dataset to facilitate their performance \cite{iglovikov2018ternausnet}.
In this paper, we would explore the potential of pre-trained embedding models (i.e., embedding models built on other sources of textual corpus or source code) in retrieving the semantics of bug reports and method code. The details are as follows.

\textbf{Semantic Retrieval of Bug Reports.}
In the file-level bug localization study by Ye et al.\cite{ye2016word}, the authors find that there is no big difference in the effectiveness of word embedding models (i.e., word2vec in their paper) trained on Wikipedia corpus or software artifacts such as API documents.
Hence we decided to directly use those well-performed pre-trained word embedding models in natural language processing area to retrieve the semantics of bug reports.
More specifically, we compared five typical/state-of-the-art pre-trained word embedding models, namely ELMo, BERT, GloVe, fastText and word2vec, in representing bug reports.
These five pre-trained word embedding models are mainly trained on Wikipedia or Google News corpus.
The dimensions of the generated word embedding vectors by five pre-trained models are different; some pre-trained word embedding models (e.g., BERT) support outputting semantic vectors of different dimensions.
In this paper, the exported dimensions of five pre-trained models are set to be 768, 768, 100, 300, and 300 for ELMo, BERT, GloVe, fastText, and word2vec, respectively.
In other words, each word of a bug report would be represented as a vector of N (i.e., the mentioned-before dimension) numeric elements. Take ELMo for example, each word of a bug report would be represented as a vector with 768 numeric values.
A bug report would be represented as a semantic matrix whose row number is its word count, and column number is the word vector dimension.
Then, we perform the max-pooling operation on each column of the matrix (i.e., selecting the maximum value of each column), to obtain the final numeric vector to represent the semantics of the whole bug report.

Before applying a pre-trained word embedding, we also do some preprocessing work on a bug report.
Specifically, we first remove stop words, symbols and numbers from each bug report.
Then we decompose each compound word into single words (e.g. ``WindowsSize'' \texttt{->} ``Windows'' + ``Size''), with the original compound word also kept.
Last, we use Porter\footnote{https://lucene.apache.org/} to stem each word and further convert each word to its lower case.
The preprocessed bug report would be taken as the input of each pre-trained word embedding model.

\textbf{Semantic Retrieval of Method Code.}
Existing studies rarely considered the structure information of method code in retrieving function semantics of methods, which negatively affects the performance of method-level bug localization techniques.
In this study, we adopt a code embedding technique called code2vec to represent the functionality of a method.
code2vec is a deep learning-based model that is built on extracted paths from the AST of methods.
By fully mining the structure information embedded in the AST paths, code2vec is able to recommend method names by given method bodies with high accuracy.
This, to a large extent, indicates that the code vectors generated by code2vec could well represent the semantics of a method.
Hence, we use code2vec to retrieve the semantics of method code in our study.
 
Similar to bug reports, we also used a pre-trained code2vec model in BLESER.
The pre-trained code2vec model is built on more than 14 million methods from the most popular 10,000 Java projects on GitHub.
The scale of the training dataset together with the diversity and popularity of Java projects makes us believe that the pre-trained model could reasonably retrieve the function semantics of methods in our experimental Java projects.
With code2vec, each method would be represented as a semantic vector of 384 numeric elements.

\subsection{Model Construction}
In this paper, we take bug localization as a learning task.
That is, we try to learn a model to predict whether a method is related to a given bug report.
We instantiate the learning task by building a convolutional neural network (CNN) model that takes the raw semantic features of bug reports and methods as instance features, with the instance label being whether a method is buggy or not for a given bug report.
During model building, we found that the training dataset is quite imbalanced, i.e., the number of buggy methods for a given bug report is generally much smaller than that of non-buggy methods.
To avoid that the model was biased by the major class, we also try to fix the imbalanced classes problem while building the CNN prediction model.
Details are as follows.

\textbf{Model Training.}
After applying pre-trained embedding models, we could obtain the semantic vectors of bug reports and method code, respectively.
These vectors are of different dimensions (e.g., a vector for a method is of 384 dimensions while a vector for a bug report with word2vec is of 300 dimensions).
In order to improve the model's learning ability and nonlinear expression ability, we use multiple fully connected layers in CNN to further extract semantic features of bug reports and code vectors.
Specifically, for a 384-dimension vector of a method, four fully connected layers are applied, i.e., \texttt{384->256->128->64->32}, where values such as 256 represent the number of nodes in each fully connected layer.
For semantic vectors of bug reports, a 768-dimension vector would be processed through 5 fully connected layers, i.e., \texttt{768->512->256->128->64->32}; a 100-dimension and 300-dimension vector would be processed through 2 and 4 fully connected layers respectively, i.e., \texttt{100->64->32} and \texttt{300->256->128->64->32}.
Following fully connected layers is the output layer. We use the logistic regression classifier as the output layer of CNN to predict the relevance between each method and a given bug report.

The whole dataset is divided into three parts, i.e., the training set, validation set, and testing set.
The training set is used to build a model, the validation set is to fine-tune the model, and the testing set is to evaluate the performance of the final model built by training and validation sets.
The features of each instance are the combination of the semantic features (i.e., embedding vectors) of a method and a bug report;
and the label of each instance is either 0 or 1, with 1 representing that the method is relevant to the bug report while 0 indicating non-relevant.
All instances are ordered by the reporting time of bug reports and would be equally partitioned into 10 parts.
The first 8 parts constitute the training set, and the 9th part and last part are taken as the validation set and testing set separately.
Three typical performance metrics are used to evaluate the effectiveness of our BLESER, including Accuracy@K, MAP and MRR (more details in Sect. \ref{perf_metrics}).

\textbf{Imbalanced Classes Handling.}
One significant problem in model training is the imbalance of class instances. That is, for a given bug report, the number of buggy methods is much smaller than the number of non-buggy methods.
As can be seen from Table \ref{info5projects}, the average number of buggy methods for a bug is about 1-2; while the total number of methods in a project is always in thousands-scale on average.
If we directly use the unbalanced dataset to train the model, the resulted model may be biased by the major class (i.e., non-buggy methods in the paper).
For example, a model may tend to predict all methods as non-relevant to achieve the best accuracy.
This is undesired in the bug localization problem as the goal of bug localization technique is to predict the truly buggy methods (i.e., the minor class) as precisely as possible.
Thus, it is essential to handle the imbalanced classes problem.
In this paper, we tested two categories of typical strategies targeting at solving the imbalanced classes problem, i.e., sampling-based strategies and cost-sensitive based strategies.

For sampling-based strategies, we tested two widely used sampling approaches, namely random under-sampling (RUS) and random over-sampling (ROS) \cite{zhou2005training}.
Random under-sampling tries to reduce the instance number of the major class to balance the classes while random over-sampling tries to enlarge the instance number of the minor class to achieve balanced classes.
More specifically, random under-sampling would eliminate the oversized major class by randomly selecting a subset of instances from the major class, with the size of the subset being equal to that of the minor class. 
Random over-sampling would expand the minor class by randomly duplicating its instances until its size is the same as that of the major class.

Unlike sampling-based strategies which try to solve the class imbalance problem by changing the distribution of class instances, cost-sensitive based strategies would not change the dataset distribution; instead, they try to balance the classes by increasing the cost of misclassified minor class instances or reducing the cost of misclassified major class instances.
In this paper, we also tested two typical cost-sensitive based strategies, namely weight-based binary-cross-entropy \cite{ho2019real} and Focal loss \cite{lin2017focal}.
The weight-based binary-cross-entropy is a weighted version of traditional binary-cross-entropy.
We use the Keras\footnote{https://github.com/keras-team/keras} package (a mainstream Python package that supports multiple deep learning frameworks) to implement this strategy, with the ``binary\_crossentropy'' function whose ``class\_weight'' is set to be ``auto''.
For the Focal loss strategy, we refer to its definition by checking the original paper that proposed it \cite{lin2017focal}, and implement it for binary classification problem in Python. The hyper parameters \begin{math} \alpha \end{math} and \begin{math} \gamma \end{math} are set to 0.25 and 2 respectively.
Both two strategies are chosen as the loss functions a model used during its training phase.

\section{Experiment setup}
\subsection{Experiment Dataset}

\begin{table*}[!t]
    \begin{center}
    \caption{Statistics of 5 experimental Java projects in Defects4J}
    \begin{tabular}{c|c|c|c|c}
        \hline
        Project Name & Bug Number & Average File Number & Average Method Number & Average Buggy Method Number\\
        \hline
        Closure compiler (Closure) & 156 & 387.69 & 5,662.25 & 1.64\\
        Apache commons-math (Math) & 85 & 497.79 & 3,326.40 & 1.38\\
        Apache commons-lang (Lang) & 56 & 89.34 & 1,868.96 & 1.23\\
        Mockito (Mockito) & 22 & 278.36 & 992.05 & 2.55\\
        Joda-Time (Time) & 23 & 156.35 & 3,099.52 & 1.96\\
        \hline
    \end{tabular}
    \label{info5projects}
    \end{center}
\end{table*}

In this paper, we used the Defects4J \footnote{https://github.com/rjust/Defects4J} dataset to evaluate our technique.
Defects4J is a publicly available dataset which is often used to evaluate bug localization and bug repair techniques  \cite{martinez2016astor, li2019deepfl, b2016learning}.
For each bug within the Defects4J dataset, the maintainers of Defects4J manually removed all code modifications that are not relevant to fixing the bug.
This guarantees that methods with modifications from Defects4J are truly buggy methods related to a bug.
The authors maintained several versions of Defects4J since its release, we used the latest version (1.5.0) by the time we conduct our experiments in our study.
The latest Defects4J dataset contains 6 software projects with 438 bugs, including JFreeChart (26 bugs), Closure compiler (176 bugs), Apache commonsmath (106 bugs), Apache commons-lang (65 bugs), Joda-Time (27 bugs) and Mockito (38 bugs).
Note that since only few bugs in JFreeChart have corresponding bug reports, we exclude JFreeChart from the experimental dataset as we are not able to obtain the semantics for those bugs whose bug reports do not exist.

Despite Defects4J provides the buggy code version and fixed code version for each bug, it does not explicitly mark the exact methods that are changed to fix a bug.
Hence, to evaluate our method-level bug localization technique, we need first to construct a dataset where each bug report is linked to its corresponding buggy methods.
We take a two-step way to help us obtain such a dataset.
Specifically, for each bug, we first use git difftool and MELD \footnote{https://meldmerge.org/} to graphically present the difference between its fixed code version and buggy code version.
Then we manually checked the code modifications, and take all methods involved in code modifications as buggy methods.
Other unchanged methods in the buggy version are regarded as non-buggy methods.

During manual check, we found that 11 bugs have no corresponding modifications to methods (i.e., these bugs reside outside of methods, e.g., in global assignment statements).
These bugs were excluded from the dataset as they are not suitable for evaluating the method-level bug localization techniques.
12 bugs residing in constructor methods, for which code2vec cannot analyze, were also excluded from the dataset.
Meanwhile, if a bug was found to link to multiple fixed commits, or share the same fixed commit with other bugs, it would also be removed; 45 bugs belonged to this category.
1 bug was found to have no corresponding deleted or modified methods but only newly added methods was also removed from the datasets as it was not applicable to predict buggy methods which did not exist yet in the buggy version of project code.
There was 1 bug from Closure compiler occurring in an inner class of an inner class in a class (i.e. multinesting). Since the Eclipse AST parser we used in our study was not able to correctly extract the buggy methods for this bug, we removed the bug in our study.
Finally, 342 bugs were left for experiments.

In this paper, we used the remained 342 bugs of 5 Java projects from the Defects4J to evaluate our BLESER.
Table \ref{info5projects} shows the basic statistics of the 5 software projects in the dataset.
We count the total number and the average number of files and methods in the buggy version of project code for all defects, by using the main directory of source code provided by Defects4J.

\subsection{Evaluation metrics}
\label{perf_metrics}
Following previous studies \cite{zhang2019finelocator, blia1.5}, We used three commonly-used metrics to evaluate our BLESER.
They are Accuracy@K, Mean Average Precision (MAP), and Mean Reciprocal Rank (MRR).

\begin{itemize}
\item Accuracy@K measures the ratio of bug reports for which at least one truly buggy method is within the top K method recommendations.
\item MAP is often used to measure an IR technology based on the mean of average precision (AvgPre) of each query in a query set \cite{manning2010introduction}. It is defined as follows:\\

    $MAP (Q) = \frac{1}{|Q|}\sum_{q \in Q}{AvgPre(q)}$\\

    $AvgPre(q)= \frac{1}{K}\sum_{k \in K}{Precision@k}$\\
    
    where {\em Q} is the query set (i.e., bug reports in this study), and {\em Precision@k} represents the precision over the top k recommendations (i.e., buggy method candidates).\\
\item MRR mainly measures the quality of the ranking results of a retrieval technology in a search task, by showing how early a relevant document (i.e., a truly buggy method) to a query (i.e., a given bug report) is retrieved \cite{voorhees1999trec}. It is defined as the mean of reciprocal rank (RR) as follows:\\
    
    $MRR (Q) = \frac{1}{|Q|}\sum_{q \in Q}{RR(q)}$\\

    $RR(q)= \frac{1}{firstRelevantPosition}$\\
\end{itemize}

\section{Experimental results}
In this section, we first present our analysis results of the effectiveness of five word embedding models and four strategies aiming at solving class imbalance problem.
Our goal is to find the most suitable word embedding model (for representing the semantics of bug reports) and the class-imbalance-problem handling strategy in the method-level bug localization task.
The most effective word embedding model and the strategy of handling class-imbalance problem would be integrated into our BLESER.
Then we would present how well our BLESER perform on five Java projects from the Defects4J dataset.

\noindent\textbf{RQ1: Does there exist a word embedding model generally outperform other models?}

Despite that a series of word embedding models have been studied much in the natural language processing field, we yet do not have a clear clue at which model we generally should refer to in retrieving the semantics of bug reports, so that we are most likely to obtain the best performance in locating bugs.
Hence, we first studied the potential of five typical word embedding models in method-level bug localization techniques, including word2vec, GloVe, fastText, ELMo, and BERT.
As shown in Fig.\ref{fig_framework}, within our bug localization framework, we used a word embedding to retrieve the semantics of bug reports, the code2vec model to retrieve method semantics, and built a CNN prediction model to locate bugs; during the CNN training process, we adopted relevant strategy to handle the class imbalance problem.

To align with the framework, we checked the potential of each word embedding model as follows:
We first tested each combination of word embedding model (5 models), code embedding model (only code2vec), and each class-imbalance handling strategy (5 strategies = 2 re-sampling strategies + 2 cost-sensitive strategies + the using-the-original-dataset strategy). In other words, we would get 5*1*5=25 combinations on each experimental project (5 projects from Defects4J).
Then, for each project by applying a given class-imbalance handling strategy, we selected the best locating results among five word embedding models in terms of MAP, MRR, and Accuracy@K.
This helps us understand the performance of each word embedding model under different class-imbalance handling strategies, and make our evaluation more general/reasonable on the whole.
After we obtained the best five results (corresponding to five class-imbalance handling strategies) for each project, we further counted the times each word embedding model occurred in the best five results.
The larger the value was, the better the word embedding was on the whole.
Table \ref{count_of_best_MAP_on_5_word_embeddings} to Table \ref{count_of_best_Accuracy_on_5_word_embeddings} show the statistic results in terms of three performance metrics, namely MAP, MRR and Accuracy@K.

\begin{table}[!t]
    \begin{center} 
    \caption{Count of best MAP from 5 word embedding models}
    \resizebox{1\linewidth}{!}{
    \begin{tabular}{c|c|c|c|c|c|c}
        \hline
        \multirow{2}{*}{\small{Word Embedding}} & \multicolumn{5}{c|}{MAP on projects} &  \multirow{2}{*}{Total} \\
        \cline{2-6}
         & Time & Mockito & Lang & Math & Closure \\
        \hline
        word2vec & 0 & 0 & 0 & 2 & 0 & 2\\
        GloVe & 2 & 0 & 2 & 1 & 0 & 5\\
        fastText & 2 & 0 & 1 & 0 & 2 & 5\\
        \textbf{ELMo} & 0 & 4 & 1 & 0 & 2 & \textbf{7} \\
        BERT & 1 & 1 & 1 & 2 & 1 & 6 \\
        \hline
    \end{tabular}}
    \label{count_of_best_MAP_on_5_word_embeddings}
    \end{center}
\end{table}

\begin{table}[!t]
    \begin{center}
    \caption{Count of best MRR from 5 word embedding models}
    \resizebox{1\linewidth}{!}{
    \begin{tabular}{c|c|c|c|c|c|c}
        \hline
        \multirow{2}{*}{\small{Word Embedding}} & \multicolumn{5}{c|}{MRR on projects} &  \multirow{2}{*}{Total} \\
        \cline{2-6}
         & Time & Mockito & Lang & Math & Closure \\
        \hline
        word2vec & 0 & 0 & 0 & 2 & 0 & 2\\
        GloVe & 2 & 0 & 2 & 1 & 2 & 7\\
        fastText & 1 & 0 & 1 & 0 & 0 & 2\\
        \textbf{ELMo} & 1 & 4 & 1 & 0 & 3 & \textbf{9} \\
        BERT & 1 & 1 & 1 & 2 & 0 & 5 \\
        \hline
    \end{tabular}}
    \label{count_of_best_MRR_on_5_word_embeddings}
    \end{center}
\end{table}

From Tab. \ref{count_of_best_MAP_on_5_word_embeddings}, we could find that among 25 best results (each project had 5 best results), ELMo is the one that occurred most frequently than other models, with 7 out of 25 best results from three projects (Mockito, Lang and Closure). The second one is BERT, with occurring in 6 best results from four projects. word2vec had the smallest number (i.e., 2) of best results.
This indicates that by using ELMo and BERT to retrieve the semantics of bug reports, is mostly likely to help us obtain the best MAP on the whole.

As related to MRR shown in Tab.\ref{count_of_best_MRR_on_5_word_embeddings}, we could also find that ELMo outperformed other models in helping achieving the best MRR, with occurring in 9 out of 25 best results from five projects.
Following ELMo is GloVe with occurring in 7 best results.
word2vec and fastText occurred least in best results (2).
This means that by using ELMo to retrieve the semantics of bug reports is mostly likely to help obtain the best MRR in locating bugs.

\begin{table*}[!t]
    \begin{center}
    \caption{Count of best Accuracy@1,5,10 from 5 word embedding models}
    \resizebox{1\linewidth}{!}{
    \begin{tabular}{c|c|c|c|c|c|c|c|c|c|c|c|c|c|c|c|c|c|c}
        \hline
        \multirow{2}{*}{Word Embedding} & \multicolumn{5}{c|}{Accuracy@1 on projects} & \multirow{2}{*}{Total} & \multicolumn{5}{c|}{Accuracy@5 on projects} & \multirow{2}{*}{Total} & \multicolumn{5}{c|}{Accuracy@10 on projects} & \multirow{2}{*}{Total}  \\
        \cline{2-6}  \cline{8-12}  \cline{14-18}
        & Time & Mockito & Lang & Math & Closure & 
        & Time & Mockito & Lang & Math & Closure &   
        & Time & Mockito & Lang & Math & Closure\\
        \hline
        word2vec & 1 & 1 & 1 & 5 & 1 & 9  & 2 & 1 & 1 & 5 & 2 & 11 & 2 & 1 & 1 & 5 & 4 & 13 \\
        GloVe & 1 & 1 & 1 & 3 & 2 & 8 & 2 & 1 & 2 & 4 & 3 & 12 & 3 & 1 & 2 & 4 & 2 & 12\\
        fastText & 2 & 1 & 1 & 3 & 1 & 8 & 2 & 1 & 1 & 5 & 3 & 12 & 2 & 1 & 1 & 5 & 2 & 11\\
        \textbf{ELMo} & 1 & 3 & 1 & 3 & 3 & \textbf{11} & 1 & 4 & 2 & 5 & 3 & \textbf{15} & 3 & 4 & 1 & 5 & 3 & \textbf{16} \\
        BERT & 1 & 0 & 1 & 4 & 2 & 8 & 2 & 1 & 2 & 4 & 3 & 12 & 2 & 1 & 2 & 5 & 2 & 12 \\
        \hline
    \end{tabular}  }
    \label{count_of_best_Accuracy_on_5_word_embeddings}
    \end{center}
\end{table*}

For the metric Accuracy@K shown in Tab. \ref{count_of_best_Accuracy_on_5_word_embeddings}, we could observe that ELMo occurred most frequently in the best results of all five Java projects, with obtaining 11, 15, and 16 best results in terms of Accuracy@1, Accuracy@5, and Accuracy@10 respectively.
This also means that ELMo could most likely help obtain the best Accuracy@K if used to retrieve the semantics of bug reports compared to other word embedding models.

\textbf{Conclusion 1:} Among five typical word embedding models, using ELMo to retrieve the semantics of bug reports could generally outperform other models in facilitating the bug localization performance.

\noindent\textbf{RQ2: Which strategy is most suitable in handling the class imbalance problem?}

To understand which strategy is most useful in helping obtain the best performance in locating bugs, we evaluated four typical class-imbalance handling strategies, including two re-sampling strategies (i.e., random over-sampling (ROS) and random under-sampling (RUS)), two cost-sensitive strategies (i.e., weight-based binary-cross entropy (WBE) and Focal Loss (Focal)).
We also tested the strategy that use the original dataset to conduct the experiments, so as to better evaluate the before-mentioned four class-imbalance handling strategies.

Specifically, similar to RQ1, after we obtained the 25 locating results (i.e., from 25 combinations of word embedding models and class-imbalance handing strategies mentioned in RQ1) for each project, we selected the best strategy among all class-imbalance handling strategies for each project under a given word embedding model in terms of MAP, MRR, and Accuracy@K.
Then we counted the times each class-imbalance handling strategy appearing in the best selected results for each project.
This helped us understand the performance of each class-imbalance handling strategy over various word embedding models, and provide a more general evaluation for each strategy.
Table \ref{count_of_best_MAP_on_5_strategies} to Table \ref{count_of_best_Accuracy_on_5_strategies} show our statistics for each class-imbalance handling strategy.

\begin{table}[!t]
    \begin{center}
    \caption{Count of best MAP under 5 class-imbalance handling strategies}
    \resizebox{1\linewidth}{!}{
    \begin{tabular}{c|c|c|c|c|c|c}
        \hline
        \multirow{2}{*}{\small{\shortstack{class-imbalance\\handling strategy}}} & \multicolumn{5}{c|}{MAP on projects} &  \multirow{2}{*}{Total} \\
        \cline{2-6}
         & Time & Mockito & Lang & Math & Closure \\
        \hline
        Original & 1 & 0 & 1 & 0 & 0 & 2\\
        \textbf{ROS} & 3 & 5 & 4 & 2 & 5 & \textbf{19}\\
        RUS & 0 & 0 & 0 & 0 & 0 & 0\\
        WBE & 1 & 0 & 0 & 2 & 0 & 3 \\
        Focal & 0 & 0 & 0 & 1 & 0 & 1 \\
        \hline
    \end{tabular}  }
    \label{count_of_best_MAP_on_5_strategies}
    \end{center}
\end{table}

\begin{table}[!t]
    \begin{center}
    \caption{Count of best MRR under 5 class-imbalance handling strategies}
    \resizebox{1\linewidth}{!}{
    \begin{tabular}{c|c|c|c|c|c|c}
        \hline
        \multirow{2}{*}{\small{\shortstack{class-imbalance\\handling strategy}}} & \multicolumn{5}{c|}{MRR on projects} &  \multirow{2}{*}{Total} \\
        \cline{2-6}
         & Time & Mockito & Lang & Math & Closure \\
        \hline
        Original & 1 & 0 & 1 & 0 & 1 & 3\\
        \textbf{ROS} & 4 & 4 & 4 & 0 & 4 & \textbf{16}\\
        RUS & 0 & 0 & 0 & 0 & 0 & 0\\
        WBE & 0 & 0 & 0 & 3 & 0 & 3 \\
        Focal & 0 & 1 & 0 & 2 & 0 & 3 \\
        \hline
    \end{tabular}  }
    \label{count_of_best_MRR_on_5_strategies}
    \end{center}
\end{table}

\begin{table*}[!t]
    \begin{center}
    \caption{Count of best Accuracy@1,5,10 under 5 class-imbalance handling strategies}
    \resizebox{1\linewidth}{!}{
    \begin{tabular}{c|c|c|c|c|c|c|c|c|c|c|c|c|c|c|c|c|c|c}
        \hline
        \multirow{2}{*}{\shortstack{class-imbalance\\handling strategy}} & \multicolumn{5}{c|}{Accuracy@1 on projects} & \multirow{2}{*}{Total} & \multicolumn{5}{c|}{Accuracy@5 on projects} & \multirow{2}{*}{Total} & \multicolumn{5}{c|}{Accuracy@10 on projects} & \multirow{2}{*}{Total}  \\
        \cline{2-6}  \cline{8-12}  \cline{14-18}
        & Time & Mockito & Lang & Math & Closure & 
        & Time & Mockito & Lang & Math & Closure &   
        & Time & Mockito & Lang & Math & Closure\\
        \hline
        Original & 1 & 1 & 1 & 3 & 3 & 9  & 2 & 1 & 2 & 5 & 1 & 11 & 4 & 1 & 3 & 5 & 1 & 14 \\
        \textbf{ROS} & 5 & 4 & 4 & 5 & 5 &  \textbf{23} & 5 & 5 & 5 & 5 & 5 &  \textbf{25} & 5 & 5 & 4 & 5 & 5 &  \textbf{24}\\
        RUS & 0 & 0 & 0 & 1 & 0 & 1 & 0 & 1 & 0 & 3 & 0 & 4 & 0 & 1 & 0 & 4 & 0 & 5\\
        WBE & 0 & 1 & 0 & 4 & 3 & 8 & 2 & 1 & 2 & 5 & 0 & 10 & 3 & 1 & 2 & 5 & 0 & 11 \\
        Focal & 0 & 0 & 0 & 5 & 0 & 5 & 0 & 0 & 0 & 5 & 0 & 5 & 0 & 0 & 0 & 5 & 0 & 5 \\
        \hline
    \end{tabular}  }
    \label{count_of_best_Accuracy_on_5_strategies}
    \end{center}
\end{table*}

From Tab.\ref{count_of_best_MAP_on_5_strategies}, we could observe that the random over-sampling (ROS) strategy outperformed far better than other strategies. Among five best combinations for each project, ROS occurred in 3, 5, 4, 2, 5 best combinations from Time to Closure project, respectively.
This means that ROS is most likely to help obtain the best MAP on the whole.

From Tab.\ref{count_of_best_MRR_on_5_strategies} and Tab.\ref{count_of_best_Accuracy_on_5_strategies}, we could observe the similar phenomenon that ROS outperformed much better than other class-imbalance handling strategies related to the MRR and Accuracy@K metrics.
Among the 25 combinations with the best MRR for five projects, the ROS occurred in 16 combinations.
Related to the Accuracy@K, ROS appeared in 23, 25, 24 out of 25 best combinations in Accuracy@1, Accuracy@5, and Accuracy@10, respectively.
This revealed that by using ROS, we are most likely to obtain the best MRR and Accuracy@K.

\textbf{Conclusion 2:} Among four typical class-imbalance handling strategies, the strategy random over-sampling could most likely help to obtain the best performance in terms of MAP, MRR and Accuracy@K on the whole.

\noindent\textbf{RQ3: How effective BLESER is in locating bugs at method level?}

Our previous two RQs has showed that ELMo outperformed other word embedding models and ROS performed much better than other class-imbalance handling strategies.
Hence, we decided to integrate ELMo and ROS, together with the code2vec code embedding model, to build our final method-level bug localization technique BLESER.
That is, ELMo was used to retrieve the semantics of bug reports, code2vec was to retrieve the semantics of method code, and the ROS was to handle the class-imbalance problem in BLESER.
Table \ref{bleserPerf} shows the locating results of BLESER on five Defects4J Java projects in terms of MAP, MRR, and Accuracy@1, Accuracy@5, and Accuracy@10.

\begin{table}[!t]
\centering
\caption{The MAP, MRR, and Accuracy@K of BLESER on five Defects4J Java projects.}
\label{bleserPerf}
\begin{tabular}{l|c|c|c|c|c}\hline
& {\bf Closure}& {\bf Math}& {\bf Lang}&{\bf Mockito}&{\bf Time}\\ \hline
MAP         &{ 0.108}   &{0.133}   &{0.226}   &{0.504} &{0.501} \\
MRR         &{ 0.155}   &{0.134}   &{0.250}   &{0.507} &{0.510} \\
Accuracy@1  &{0.133}    &{0.125}   &{0.2}     &{0.5}   &{0.5}   \\
Accuracy@5  &{ 0.133}   &{0.125}   &{0.4}     &{0.5}   &{0.5}   \\
Accuracy@10 &{ 0.133}   &{0.125}   &{0.4}     &{0.5}   &{0.5}   \\ \hline
\end{tabular}
\end{table}

From the table, we can find that
BLESER can achieve MAP of 0.108, 0.133,0.226, 0.504, and 0.501 on Closure, Math, Lang, Mockito, and Time respectively. The best MAP were obtained on Mockito and Time with MAP larger than 0.5.
While for MRR, BLESER achieved MRR from 0.134 to 0.510 on five projects, with the worst performance 0.134 on Math and the best performance 0.510 on Time project.
As for the Accuracy@K performance, we can observe that when K=1, BLESER could recommend buggy methods correctly for 13.3\%, 12.5\%, 20\%, 50\%, and 50\% bug reports on Closure, Math, Lang, Mockito, and Time, respectively.
Meanwhile, when K increase (e.g., to 5 or 10), except the project Lang with accuracy increasing from 20\% to 40\%, the accuracy@K values remain the same as Accuracy@1.
This indicates that BLESER could on one hand relatively easily help developers find the first buggy method for a given bug report, on the hand may still have some limitations in finding the other buggy methods.
However, considering that each bug report generally have less than two buggy methods (as statistics in Table \ref{info5projects}), BLESER is still promising in the task of method-level bug localization.
It would be valuable to explore the way to better retrieve other remaining buggy methods after we retrieve the first buggy method with BLESER for a new coming bug report with multiple buggy methods.

\textbf{Conclusion 3:} BLESER could achieve MAP of 0.108-0.504, MRR of 0.134-0.510, and Accuracy@1 of 0.125-0.5 on five Defects4J projects.

\section{Threats to validity}
\textbf{Internal Validity.}
In this paper, we directly used the default parameter settings of pre-trained embedding models during semantic retrieving for bug localization task.
We have to admit that the difference in domains (i.e., bug localization vs. other natural language processing or software engineering tasks) theoretically require customized parameter settings so as to obtain the best performance.
However, given that default settings generally reflect the best practice or valuable experience of researchers and tool developers, we believe that it is still reasonable and acceptable to adopt the default parameter settings of pre-trained models.
We would try to explore how different parameter settings affect method-level bug localization performance in the future.

\textbf{External Validity.} 
Currently, we only demonstrated the effectiveness of our approach on 5 Java projects from the Defects4J dataset. We cannot guarantee that our arrived conclusions are applicable to other industrial or open source projects.
Nevertheless, the adoption of (pre-trained) word embedding and code embedding techniques within our approach still shed some light on the resolution of e.g., cross-project bug localization problems. In the future, we would try to evaluate our approach on more and larger projects to improve the generalization of the conclusions in this paper.

\section{Related Work}
Researchers have done a series of work to automatically locate bugs at method granularity.
These studies could be roughly divided into three categories: static bug localization techniques, dynamic bug localization techniques and hybrid bug localization techniques which combine both static and dynamic techniques.
\subsection{static method-level bug localization techniques}
Static techniques mainly make use of bug reports, source code and other static artifacts (e.g., commit logs) generated in the development process to locate bugs \cite{zhang2019finelocator}.
The basic idea of static method-level bug localization techniques is to find suspicious methods related to a bug report by extracting some semantic features from bug reports and methods (sometimes also include other static artifacts), and matching them by e.g., calculating their textual similarities.
A main-stream category of such techniques are those based on information retrieval technologies (such as vector space model (VSM), latent semantic indexing (LSI), latent dirichlet allocation (LDA), etc.) \cite{scanniello2015link, corley2015modeling, sun2015msr4sm, zhang2016inferring, zhang2018fusing, eddy2018impact}.

Poshyvanyk et al. first explored the potential of LSI model for bug localization, and then proposed a bug localization approach that combined formal concept analysis and LSI \cite{poshyvanyk2007combining}. 
Dit et al. found that more accurate word segmentation preprocessing technologies could facilitate LSI-based bug localization techniques \cite{Dit2011Can}.
Gay et al. proposed to use relevance feedback to reconstruct the bug report query to improve a VSM-based bug localization technique  \cite{gay2009use}.
Scanniello et al. proposed to use PageRank algorithm to extract the dependencies of methods to also facilitate a VSM-based bug localization model \cite{scanniello2015link}.
Davies et al. proposed to use historically similar bug reports to improve the TF-IDF bug localization approach \cite{davies2012using}.
Lukins et al. proposed to use the LDA topic model to predict suspicious methods for a given bug report \cite{lukins2008source, lukins2010bug}. 
Biggers et al. studied the effects of different confituration settings on the performance of LDA-based bug localization techniques \cite{biggers2014configuring}. 
To avoid repeated training, Corley et al. proposed a LDA-based bug localization technique for code modification increment \cite{corley2015modeling}.
Sun et al. proposed to use LDA to retrieve relevant information from data sources such as version control systems, bug report repositories and email archives to improve existing bug localization techniques \cite{sun2015msr4sm}.
In order to solve the problem that a single topic model may fail to extract higher-level semantics, Zhang et al. proposed a bug localization technique with multi-abstraction vector space model which uses the LDA model to represent bug reports and methods into several abstraction levels of topics\cite{zhang2016inferring, zhang2018fusing}. 
Eddy et al. analyzed the effect of using different weighting strategies on elements such as function annotations and local variables on the performance of LDA-based bug localization techniques \cite{eddy2018impact}.
Dit et al. did a literature review on existing feature (bug) localization techniques \cite{dit2013feature}. 
Razzaq et al. compared eight bug localization techniques based on VSM, LSI and LDA, and found that the effectiveness of these techniques would be affected by the characteristics of the datasets and the applied techniques \cite{razzaq2020empirical}.

The above research mainly explored the potential of traditional information retrieval techniques in bug localization.
There also exist some studies that try to improve existing bug localization techniques from other perspectives (such as using method call dependency, etc.). 
Youm et al. tried to locate suspicious source code files at first and then locate suspicious methods in the top-10 suspicious files. They make use of various data sources including bug reports, code change histories, stacktraces from bug reports \cite{youm2017improved}.
Zhang et al. proposed to use the semantic similarity, the recency of modification time and the call dependency between methods to augment the vector representations of methods to improve method-level bug localization \cite{zhang2019finelocator}. 
Chochlov et al. added relevant commit logs to method bodies and take the expand methods as to-be-queried document corpus for a bug report \cite{chochlov2017historical}. 
Shu et al. developed a causal inference based bug localization technique to reduce confounding bias \cite{shu2013mfl}.

Our BLESER was also a static method-level bug localization technique.
Unlike above static techniques, we took full consideration of code structure information of methods in retrieving semantic features of method code. We further investigated how typical/state-of-the-art word embedding techniques could help in representing bug reports and facilitate bug localization techniques. 
 
\subsection{Dynamic method-level bug localization techniques}
Dynamic bug localization techniques usually need to run the to-be-located software.
They mainly rely on analyzing program running information collected through code instrumentation, execution monitoring and formal analysis to locate suspicious methods. 
Spectrum-based fault localization (SBFL) techniques are representatives of dynamic bug localization techniques \cite{xuan2014learning, b2016learning, sohn2019empirical, sohn2019train, zhang2017boosting, laghari2016fine, laghari2018use, lou2019can}.

A SBFL technique generally use the coverage information of program entities (methods in this study) executed in failed and passed test suites to measure the suspiciousness of each program entity \cite{xuan2014learning}.
The basic idea of SBFL techniques is that program entities executed by failed test cases are more suspicious than those primarily executed by passed test cases.
Till now, researchers have designed various suspiciousness formula based on coverage information to locate bugs, including Tarantula, Wong1, etc. \cite{yoo2017human}.
Xuan et al. proposed a learning-to-rank method to combine multiple suspiciousness formulas to obtain better localization performance than single formula alone\cite{xuan2014learning}. 
Le et al. also used a learning-to-rank method to combine suspiciousness formulas and invariants inferred from executions of failed/passed test cases to locate most suspicious methods \cite{b2016learning}. 
Sohn et al. proposed to take SBFL suspiciousness formulas and the characteristics of source code and code churns as features and applied genetic programming (GP) to these features to locate bugs \cite{sohn2019empirical}.
Unlike existing studies always choosing the model with best performance from a series of training models, Sohn et al. designed an integration strategy to combine all training models to achieve better localization performance by exploiting the advantages of individual models \cite{sohn2019train}.
Zhang et al. used the PageRank algorithm to differentiate test cases to recalculate the spectrum information, and then used this information to calculate the suspiciousness of methods \cite{zhang2017boosting}.
Laghari et al. proposed to use frequency itemset mining to obtain the internal method invocation pattern of each method to improve the effectiveness of SBFL techniques \cite{laghari2016fine, laghari2018use}. 
Lou et al. explored the possibility of using bug repair techniques to facilitate SBFL techniques \cite{lou2019can}.

In addition to the above (improved) SBFL techniques, some researchers also developed some mutation-based bug localization techniques \cite{musco2016mutation, de2018mutation}.
Musco et al. proposed a mutation-based graph inference approach to locate bugs \cite{musco2016mutation}.
They first performed a series of tests on a bunch of program mutants to build a method call graph.
then, the Killed mutants and relevant failed test cases were used to construct a causal diagram.
When a program failed a test case, they used the constructed causal graph and traditional SBFL techniques to find the suspicious methods.
De-Freitas et al. used a GP technology to evolve mutation formulas to obtain better mutation formulas in locating bugs \cite{de2018mutation}.
In this study, we mainly focus on developing a static method-level bug localization in attempting to fully extract semantic features of bug reports and method code without running software programs.

\subsection{Hybrid method-level bug localization techniques}
In order to make full use of the advantages of dynamic and static techniques, some researchers developed hybrid techniques which combined both dynamic and static techniques in locating bugs\cite{poshyvanyk2007feature, le2015information, dao2017does, hoang2018network, li2019deepfl}.
Poshyvanyk et al. combined a probabilistic ranking model of methods based on execution scenarios and a LSI-based model to perform bug localization \cite{poshyvanyk2007feature}.
Le et al. proposed the AML framework that combined an information retrieval based technique and a SBFL technique to predict  suspicious methods for a given bug \cite{le2015information}. 
Dao et al. used dynamic execution information to improve the effectiveness of information retrieval based bug localization techniques \cite{dao2017does}. 
Hoang et al. proposed a network-clustered multi-modal bug localization technique \cite{hoang2018network}.
This technique tried to locate bugs by integrating static information contained in a bug report and dynamic information from program spectrums.
Li et al. used deep learning models to combine various kinds of information to locate bugs, including code complexity, suspicious scores calculated from spectrum information and program mutation information \cite{li2019deepfl}.
Unlike above studies, we mainly focus on exploiting static information embedded in bug reports and method code to locate bugs.
\section*{Conclusion}
In this paper, we proposed a static bug localization technique called BLESER to locate bugs at method granularity.
Within BLESER, the semantics of methods and bug reports were retrieved by an AST-based code embedding model
and a word embedding model, respectively;
and a CNN was built to leverage these two kinds of semantic features to predict whether a method is buggy or not for a given bug report.
To understand the potential of word embedding models in facilitating method-level bug localization, We compared five traditional/state-of-the-art embedding techniques, including word2vec, fastText, GloVe, ELMo, and BERT.
We found that ELMo could help obtain the best performance than other four models on the whole.
We also attempted to address the class imbalance problem while constructing the CNN model.
We found that the random over-sampling strategy outperformed the other three strategies in handling class imbalance problem in the bug localization task.
By evaluating BLESER (integrating ELMo, code2vec, and ROS) on the 5 Java projects from the Defects4J dataset, we found that
BLESER could achieve MAP of 0.108-0.504, MRR of 0.134-0.510, and Accuracy@1 of 0.125-0.5 on five Defects4J projects.

\textbf{Future Work.}
Currently, we mainly made use of two data sources, i.e., bug reports and method code, to do bug localization. In the future, we plan to study how to effectively integrate other (possibly) useful data sources such as identifiers and commit logs to improve BLESER.
Besides, we also plan to do an empirical study on those bugs for which our BLESER failed to recommend truly buggy methods, in the hope that we could find more useful hints in improving our method-level bug localization technique.

\bibliographystyle{IEEEtran}
\bibliography{reference}

\begin{thebibliography}{10}
\providecommand{\url}[1]{#1}
\csname url@samestyle\endcsname
\providecommand{\newblock}{\relax}
\providecommand{\bibinfo}[2]{#2}
\providecommand{\BIBentrySTDinterwordspacing}{\spaceskip=0pt\relax}
\providecommand{\BIBentryALTinterwordstretchfactor}{4}
\providecommand{\BIBentryALTinterwordspacing}{\spaceskip=\fontdimen2\font plus
\BIBentryALTinterwordstretchfactor\fontdimen3\font minus
  \fontdimen4\font\relax}
\providecommand{\BIBforeignlanguage}[2]{{%
\expandafter\ifx\csname l@#1\endcsname\relax
\typeout{** WARNING: IEEEtran.bst: No hyphenation pattern has been}%
\typeout{** loaded for the language `#1'. Using the pattern for}%
\typeout{** the default language instead.}%
\else
\language=\csname l@#1\endcsname
\fi
#2}}
\providecommand{\BIBdecl}{\relax}
\BIBdecl

\bibitem{kochhar2016practitioners}
P.~S. Kochhar, X.~Xia, D.~Lo, and S.~Li, ``Practitioners' expectations on
  automated fault localization,'' in \emph{Proceedings of the 25th
  International Symposium on Software Testing and Analysis}, 2016, pp.
  165--176.

\bibitem{zou2018practitioners}
W.~Zou, D.~Lo, Z.~Chen, X.~Xia, Y.~Feng, and B.~Xu, ``How practitioners
  perceive automated bug report management techniques,'' \emph{IEEE
  Transactions on Software Engineering}, 2018.

\bibitem{zhang2019finelocator}
W.~Zhang, Z.~Li, Q.~Wang, and J.~Li, ``Finelocator: A novel approach to
  method-level fine-grained bug localization by query expansion,''
  \emph{Information and Software Technology}, vol. 110, pp. 121--135, 2019.

\bibitem{lukins2010bug}
S.~K. Lukins, N.~A. Kraft, and L.~H. Etzkorn, ``Bug localization using latent
  dirichlet allocation,'' \emph{Information and Software Technology}, vol.~52,
  no.~9, pp. 972--990, 2010.

\bibitem{zhang2016inferring}
Y.~Zhang, D.~Lo, X.~Xia, T.-D.~B. Le, G.~Scanniello, and J.~Sun, ``Inferring
  links between concerns and methods with multi-abstraction vector space
  model,'' in \emph{Proceedings of the International Conference on Software
  Maintenance and Evolution}, 2016, pp. 110--121.

\bibitem{li2018bridging}
X.~Li, H.~Jiang, Y.~Kamei, and X.~Chen, ``Bridging semantic gaps between
  natural languages and apis with word embedding,'' \emph{IEEE Transactions on
  Software Engineering}, 2018.

\bibitem{ye2016word}
X.~Ye, H.~Shen, X.~Ma, R.~Bunescu, and C.~Liu, ``From word embeddings to
  document similarities for improved information retrieval in software
  engineering,'' in \emph{Proceedings of the 38th international conference on
  software engineering}, 2016, pp. 404--415.

\bibitem{b2016learning}
T.-D. B.~Le, D.~Lo, C.~Le~Goues, and L.~Grunske, ``A learning-to-rank based
  fault localization approach using likely invariants,'' in \emph{Proceedings
  of the 25th International Symposium on Software Testing and Analysis}, 2016,
  pp. 177--188.

\bibitem{li2019deepfl}
X.~Li, W.~Li, Y.~Zhang, and L.~Zhang, ``Deepfl: Integrating multiple fault
  diagnosis dimensions for deep fault localization,'' in \emph{Proceedings of
  the 28th ACM SIGSOFT International Symposium on Software Testing and
  Analysis}, 2019, pp. 169--180.

\bibitem{martinez2016astor}
M.~Martinez and M.~Monperrus, ``Astor: A program repair library for java,'' in
  \emph{Proceedings of the 25th International Symposium on Software Testing and
  Analysis}, 2016, pp. 441--444.

\bibitem{harris1954distributional}
Z.~Harris, ``Distributional structure. word, 10 (2-3): 146--162. reprinted in
  fodor, j. a and katz, jj (eds.), readings in the philosophy of language,''
  1954.

\bibitem{collobert2008unified}
R.~Collobert and J.~Weston, ``A unified architecture for natural language
  processing: Deep neural networks with multitask learning,'' in
  \emph{Proceedings of the 25th International Conference on Machine Learning},
  2008, pp. 160--167.

\bibitem{mikolov2013distributed}
T.~Mikolov, I.~Sutskever, K.~Chen, G.~S. Corrado, and J.~Dean, ``Distributed
  representations of words and phrases and their compositionality,'' in
  \emph{Proceedings of the Advances in neural information processing systems},
  2013, pp. 3111--3119.

\bibitem{chen2016mining}
C.~Chen, S.~Gao, and Z.~Xing, ``Mining analogical libraries in q\&a
  discussions--incorporating relational and categorical knowledge into word
  embedding,'' in \emph{Proceedings of the 23rd International Conference on
  Software Analysis, evolution, and reengineering}, vol.~1, 2016, pp. 338--348.

\bibitem{alon2019code2vec}
U.~Alon, M.~Zilberstein, O.~Levy, and E.~Yahav, ``code2vec: Learning
  distributed representations of code,'' \emph{Proceedings of the ACM on
  Programming Languages}, vol.~3, no. POPL, pp. 1--29, 2019.

\bibitem{zhang2019novel}
J.~Zhang, X.~Wang, H.~Zhang, H.~Sun, K.~Wang, and X.~Liu, ``A novel neural
  source code representation based on abstract syntax tree,'' in
  \emph{Proceedings of the IEEE/ACM 41st International Conference on Software
  Engineering}, 2019, pp. 783--794.

\bibitem{white2016deep}
M.~White, M.~Tufano, C.~Vendome, and D.~Poshyvanyk, ``Deep learning code
  fragments for code clone detection,'' in \emph{Proceedings of the 31st
  International Conference on Automated Software Engineering}, 2016, pp.
  87--98.

\bibitem{mou2016convolutional}
L.~Mou, G.~Li, L.~Zhang, T.~Wang, and Z.~Jin, ``Convolutional neural networks
  over tree structures for programming language processing,'' in
  \emph{Proceedings of the 30th AAAI Conference on Artificial Intelligence},
  2016.

\bibitem{socher2011parsing}
R.~Socher, C.~C. Lin, C.~Manning, and A.~Y. Ng, ``Parsing natural scenes and
  natural language with recursive neural networks,'' in \emph{Proceedings of
  the 28th International Conference on Machine Learning}, 2011, pp. 129--136.

\bibitem{tai2015improved}
K.~S. Tai, R.~Socher, and C.~D. Manning, ``Improved semantic representations
  from tree-structured long short-term memory networks,'' \emph{arXiv preprint
  arXiv:1503.00075}, 2015.

\bibitem{peters2018deep}
M.~E. Peters, M.~Neumann, M.~Iyyer, M.~Gardner, C.~Clark, K.~Lee, and
  L.~Zettlemoyer, ``Deep contextualized word representations,'' \emph{arXiv
  preprint arXiv:1802.05365}, 2018.

\bibitem{wang2016cnn}
J.~Wang, Y.~Yang, J.~Mao, Z.~Huang, C.~Huang, and W.~Xu, ``Cnn-rnn: A unified
  framework for multi-label image classification,'' in \emph{Proceedings of the
  29th IEEE conference on computer vision and pattern recognition}, 2016, pp.
  2285--2294.

\bibitem{mikolov2017advances}
T.~Mikolov, E.~Grave, P.~Bojanowski, C.~Puhrsch, and A.~Joulin, ``Advances in
  pre-training distributed word representations,'' \emph{arXiv preprint
  arXiv:1712.09405}, 2017.

\bibitem{iglovikov2018ternausnet}
V.~Iglovikov and A.~Shvets, ``Ternausnet: U-net with vgg11 encoder pre-trained
  on imagenet for image segmentation,'' \emph{arXiv preprint arXiv:1801.05746},
  2018.

\bibitem{zhou2005training}
Z.~Zhou and X.~Liu, ``Training cost-sensitive neural networks with methods
  addressing the class imbalance problem,'' \emph{IEEE Transactions on
  knowledge and data engineering}, vol.~18, no.~1, pp. 63--77, 2005.

\bibitem{ho2019real}
Y.~Ho and S.~Wookey, ``The real-world-weight cross-entropy loss function:
  Modeling the costs of mislabeling,'' \emph{IEEE Access}, vol.~8, pp.
  4806--4813, 2019.

\bibitem{lin2017focal}
T.-Y. Lin, P.~Goyal, R.~Girshick, K.~He, and P.~Doll{\'a}r, ``Focal loss for
  dense object detection,'' in \emph{Proceedings of the IEEE international
  conference on computer vision}, 2017, pp. 2980--2988.

\bibitem{blia1.5}
K.~C. Youm, J.~Ahn, and E.~Lee, ``Improved bug localization based on code
  change histories and bug reports,'' \emph{Information and Software
  Technology}, vol.~82, pp. 177--192, 2017.

\bibitem{manning2010introduction}
C.~Manning, P.~Raghavan, and H.~Sch{\"u}tze, ``Introduction to information
  retrieval,'' \emph{Natural Language Engineering}, vol.~16, no.~1, pp.
  100--103, 2010.

\bibitem{voorhees1999trec}
E.~M. Voorhees \emph{et~al.}, ``The trec-8 question answering track report,''
  vol.~99, pp. 77--82, 1999.

\bibitem{scanniello2015link}
G.~Scanniello, A.~Marcus, and D.~Pascale, ``Link analysis algorithms for static
  concept location: an empirical assessment,'' \emph{Empirical Software
  Engineering}, vol.~20, no.~6, pp. 1666--1720, 2015.

\bibitem{corley2015modeling}
C.~S. Corley, K.~L. Kashuda, and N.~A. Kraft, ``Modeling changeset topics for
  feature location,'' in \emph{Proceedings of the International Conference on
  Software Maintenance and Evolution}, 2015, pp. 71--80.

\bibitem{sun2015msr4sm}
X.~Sun, B.~Li, H.~Leung, B.~Li, and Y.~Li, ``Msr4sm: Using topic models to
  effectively mining software repositories for software maintenance tasks,''
  \emph{Information and Software Technology}, vol.~66, pp. 1--12, 2015.

\bibitem{zhang2018fusing}
Y.~Zhang, D.~Lo, X.~Xia, G.~Scanniello, T.-D.~B. Le, and J.~Sun, ``Fusing
  multi-abstraction vector space models for concern localization,''
  \emph{Empirical Software Engineering}, vol.~23, no.~4, pp. 2279--2322, 2018.

\bibitem{eddy2018impact}
B.~P. Eddy, N.~A. Kraft, and J.~Gray, ``Impact of structural weighting on a
  latent dirichlet allocation--based feature location technique,''
  \emph{Journal of Software: Evolution and Process}, vol.~30, no.~1, p. e1892,
  2018.

\bibitem{poshyvanyk2007combining}
D.~Poshyvanyk and A.~Marcus, ``Combining formal concept analysis with
  information retrieval for concept location in source code,'' in
  \emph{Proceedings of the 15th International Conference on Program
  Comprehension}, 2007, pp. 37--48.

\bibitem{Dit2011Can}
B.~Dit, L.~Guerrouj, D.~Poshyvanyk, and G.~Antoniol, ``Can better identifier
  splitting techniques help feature location?'' in \emph{Proceedings of the
  19th International Conference on Program Comprehension}, 2011, pp. 11--20.

\bibitem{gay2009use}
G.~Gay, S.~Haiduc, A.~Marcus, and T.~Menzies, ``On the use of relevance
  feedback in ir-based concept location,'' in \emph{Proceedings of the
  International Conference on Software Maintenance}, 2009, pp. 351--360.

\bibitem{davies2012using}
S.~Davies, M.~Roper, and M.~Wood, ``Using bug report similarity to enhance bug
  localisation,'' in \emph{Proceedings of the 19th Working Conference on
  Reverse Engineering}, 2012, pp. 125--134.

\bibitem{lukins2008source}
S.~K. Lukins, N.~A. Kraft, and L.~H. Etzkorn, ``Source code retrieval for bug
  localization using latent dirichlet allocation,'' in \emph{Proceedings of the
  15th Working Conference on Reverse Engineering}, 2008, pp. 155--164.

\bibitem{biggers2014configuring}
L.~R. Biggers, C.~Bocovich, R.~Capshaw, B.~P. Eddy, L.~H. Etzkorn, and N.~A.
  Kraft, ``Configuring latent dirichlet allocation based feature location,''
  \emph{Empirical Software Engineering}, vol.~19, no.~3, pp. 465--500, 2014.

\bibitem{dit2013feature}
B.~Dit, M.~Revelle, M.~Gethers, and D.~Poshyvanyk, ``Feature location in source
  code: a taxonomy and survey,'' \emph{Journal of software: Evolution and
  Process}, vol.~25, no.~1, pp. 53--95, 2013.

\bibitem{razzaq2020empirical}
A.~Razzaq, A.~Le~Gear, C.~Exton, and J.~Buckley, ``An empirical assessment of
  baseline feature location techniques,'' \emph{Empirical Software
  Engineering}, vol.~25, no.~1, pp. 266--321, 2020.

\bibitem{youm2017improved}
K.~C. Youm, J.~Ahn, and E.~Lee, ``Improved bug localization based on code
  change histories and bug reports,'' \emph{Information and Software
  Technology}, vol.~82, pp. 177--192, 2017.

\bibitem{chochlov2017historical}
M.~Chochlov, M.~English, and J.~Buckley, ``A historical, textual analysis
  approach to feature location,'' \emph{Information and Software Technology},
  vol.~88, pp. 110--126, 2017.

\bibitem{shu2013mfl}
G.~Shu, B.~Sun, A.~Podgurski, and F.~Cao, ``Mfl: Method-level fault
  localization with causal inference,'' in \emph{Proceedings of the 6th
  International Conference on Software Testing, Verification and Validation},
  2013, pp. 124--133.

\bibitem{xuan2014learning}
J.~Xuan and M.~Monperrus, ``Learning to combine multiple ranking metrics for
  fault localization,'' in \emph{Proceedings of the International Conference on
  Software Maintenance and Evolution}, 2014, pp. 191--200.

\bibitem{sohn2019empirical}
J.~Sohn and S.~Yoo, ``Empirical evaluation of fault localisation using code and
  change metrics,'' \emph{IEEE Transactions on Software Engineering}, 2019.

\bibitem{sohn2019train}
------, ``Why train-and-select when you can use them all? ensemble model for
  fault localisation,'' in \emph{Proceedings of the Genetic and Evolutionary
  Computation Conference}, 2019, pp. 1408--1416.

\bibitem{zhang2017boosting}
M.~Zhang, X.~Li, L.~Zhang, and S.~Khurshid, ``Boosting spectrum-based fault
  localization using pagerank,'' in \emph{Proceedings of the 26th ACM SIGSOFT
  International Symposium on Software Testing and Analysis}, 2017, pp.
  261--272.

\bibitem{laghari2016fine}
G.~Laghari, A.~Murgia, and S.~Demeyer, ``Fine-tuning spectrum based fault
  localisation with frequent method item sets,'' in \emph{Proceedings of the
  31st IEEE/ACM International Conference on Automated Software Engineering},
  2016, pp. 274--285.

\bibitem{laghari2018use}
G.~Laghari and S.~Demeyer, ``On the use of sequence mining within spectrum
  based fault localisation,'' in \emph{Proceedings of the 33rd Annual ACM
  Symposium on Applied Computing}, 2018, pp. 1916--1924.

\bibitem{lou2019can}
Y.~Lou, A.~Ghanbari, X.~Li, L.~Zhang, D.~Hao, and L.~Zhang, ``Can automated
  program repair refine fault localization?'' \emph{arXiv preprint
  arXiv:1910.01270}, 2019.

\bibitem{yoo2017human}
S.~Yoo, X.~Xie, F.-C. Kuo, T.~Y. Chen, and M.~Harman, ``Human competitiveness
  of genetic programming in spectrum-based fault localisation: Theoretical and
  empirical analysis,'' \emph{ACM Transactions on Software Engineering and
  Methodology (TOSEM)}, vol.~26, no.~1, pp. 1--30, 2017.

\bibitem{musco2016mutation}
V.~Musco, M.~Monperrus, and P.~Preux, ``Mutation-based graph inference for
  fault localization,'' in \emph{Proceedings of the 16th International Working
  Conference on Source Code Analysis and Manipulation (SCAM)}, 2016, pp.
  97--106.

\bibitem{de2018mutation}
D.~M. De-Freitas, P.~S. Leitao-Junior, C.~G. Camilo-Junior, and R.~Harrison,
  ``Mutation-based evolutionary fault localisation,'' in \emph{Proceedings of
  the Congress on Evolutionary Computation}, 2018, pp. 1--8.

\bibitem{poshyvanyk2007feature}
D.~Poshyvanyk, Y.-G. Gueheneuc, A.~Marcus, G.~Antoniol, and V.~Rajlich,
  ``Feature location using probabilistic ranking of methods based on execution
  scenarios and information retrieval,'' \emph{Transactions on Software
  Engineering}, vol.~33, no.~6, pp. 420--432, 2007.

\bibitem{le2015information}
T.-D.~B. Le, R.~J. Oentaryo, and D.~Lo, ``Information retrieval and spectrum
  based bug localization: better together,'' in \emph{Proceedings of the 10th
  Joint Meeting on Foundations of Software Engineering}, 2015, pp. 579--590.

\bibitem{dao2017does}
T.~Dao, L.~Zhang, and N.~Meng, ``How does execution information help with
  information-retrieval based bug localization?'' in \emph{Proceedings of the
  IEEE/ACM 25th International Conference on Program Comprehension (ICPC)},
  2017, pp. 241--250.

\bibitem{hoang2018network}
T.~Hoang, R.~J. Oentaryo, T.-D.~B. Le, and D.~Lo, ``Network-clustered
  multi-modal bug localization,'' \emph{IEEE Transactions on Software
  Engineering}, vol.~45, no.~10, pp. 1002--1023, 2018.

\end{thebibliography}

\end{document}